\documentclass[sigconf]{acmart}
\usepackage{subcaption}
\usepackage{graphicx}
\usepackage{array}
\usepackage{xcolor}

\AtBeginDocument{%
  }

\setcopyright{acmlicensed}
\copyrightyear{2025}
\acmYear{2025}
\setcopyright{cc}
\setcctype{by}
\acmConference[Koli Calling '25]{25th Koli Calling International Conference on Computing Education Research}{November 11--16, 2025}{Koli, Finland}
\acmBooktitle{25th Koli Calling International Conference on Computing Education Research (Koli Calling '25), November 11--16, 2025, Koli, Finland}
\acmPrice{}
\acmDOI{10.1145/3769994.3770037}
\acmISBN{979-8-4007-1599-0/25/11}

\begin{document}

\title{“Grillz on a hijabi”: Intersectional Identities in Fostering Critical AI Literacy}


\author{Jaemarie Solyst}
\affiliation{%
  \institution{University of Washington}
  \city{Seattle}
  \country{USA}}
\email{jaemarie@cs.washington.edu}

\author{Chloe Fong}
\affiliation{%
  \institution{University of Washington}
  \city{Seattle}
  \country{USA}}
\email{cf2024@cs.washington.edu}

\author{Faisal Nurdin}
\affiliation{%
  \institution{University of Washington}
  \city{Seattle}
  \country{USA}}
\email{fmnurdin@uw.edu}

\author{Rotem Landesman}
\affiliation{%
  \institution{University of Washington}
  \city{Seattle}
  \country{USA}}
\email{roteml@uw.edu}

\author{R. Benjamin Shapiro}
\affiliation{%
  \institution{University of Washington}
  \city{Seattle}
  \country{USA}}
\email{rbs@cs.washington.edu}

\renewcommand{\shortauthors}{Solyst et al.}

\begin{abstract}
As AI increasingly saturates our daily lives, it is crucial that youth develop skills to critically use and assess AI systems and envision better alternatives. We apply theories from culturally responsive computing to design and study a learning experience meant to support Black Muslim teen girls in developing critical literacy with generative AI (GenAI). We investigate fashion design as a culturally-rich, creative domain for youth to apply GenAI and then reflect on GenAI’s socio-ethical aspects in relation to their own intersectional identities. Through a case study of a three-day, voluntary informal education program, we show how fashion design with GenAI exposed  affordances and limitations of current GenAI tools. As the girls used GenAI to create realistic depictions of their dream fashion collections, they encountered socio-ethical limitations of AI, such as biased models and malfunctioning safety systems that prohibited their generation of outputs that reflected their creative ideas, bodies, and cultures. Discussions anchored in the phenomenology of impossible creative realization  supported participants’ development of critical AI literacy and descriptions of how preferable, identity-affirming technologies would behave. Our findings contribute to the field’s growing understanding of how computing education experience designs linking creativity and identity can support critical AI literacy development.
\end{abstract}




\maketitle

\section{Introduction}
As artificial intelligence (AI) becomes increasingly embedded in everyday life, youth are engaging with AI technologies more frequently. From AI-driven recommendation systems (e.g.,  TikTok feeds) to generative AI (GenAI) chatbots, AI impacts how young people learn, reflect, and communicate. However, AI has been well-documented to cause harm, such as through the reproduction and amplification of problematic societal biases \cite{shelby2023sociotechnical}. This is especially the case for systemically marginalized communities—Black users and girls, for example, have faced disproportionately high rates of harms from AI systems that stereotype, erase, and exclude them (e.g., \cite{epps2020racism, rittenhouse2022algorithms}). In the United States, where we conducted this research, these harms add to a long history of violence toward Black people, originating in the slave trade \cite{coates2015case} and continuing through today, when the maternal death rates of Black women are triple those of White women and the gap between them is widening \cite{hoyert2025health}. Challenging oppressive systems requires understanding them, so we center Black teen girls learning about AI in this work. While fixing AI issues is  not the responsibility of those impacted by them, critical AI literacy can support navigation and interrogation of such problematic, ubiquitous systems in the age of AI. Further, in order for a more just future of AI, design decisions and assessment need to include those currently pushed to the margins of technology development---critical AI literacy can empower greater engagement with socio-ethical topics of AI  \cite{solyst2025rad}.

Since roughly 2017, efforts to foster AI literacy for teens have focused on supporting their use and/or creation of ML models   (e.g., curating small training datasets to finetune custom models that could be deployed in youth-made software \cite{druga2018growing,zimmermann2019youth}). However, recent work has put more emphasis on exploring the critical, socio-ethical aspects of AI literacy. By critical AI literacy, we mean the ability to reflect on systems of power embedded in AI, as well as the limitations and socio-ethical implications of AI \cite{Long2020, solyst2025rad}. Prior work has found that youth make the most sense of socio-ethical aspects of AI when they can connect topics to their lived experiences and funds of knowledge \cite{salac2023funds, solyst2025investigating, solyst2023potential, solyst2023would, register2020learning}. However, this prior work uses somewhat closed-ended approaches or heavy scaffolding, despite literature suggesting that more open-ended efforts may support deeper sensemaking; more open-ended designs better support students bringing their own experiences and interests into learning experiences \cite{lee2003toward}. Additionally, despite these efforts, there is still a great need to better engage systemically marginalized youth, and to better understand \textbf{how} aspects of identity are resources for fostering AI literacy. Developing these  insights will support future  design of AI literacy learning experiences that can center a diverse range of youth.

In this work, we draw on prior research on culturally responsive pedagogy in computing \cite{scott2015culturally}, and on cultural modeling \cite{lee2003toward}, which have been successful in meaningfully centering Black girls in STEM and computing education initiatives, in addition to traditional literacies. Such efforts carefully build upon learners’ identities and cultures, including their everyday, family- and community-connected cultural practices. Working in collaboration with a non-profit organization that primarily exists to support flourishing within a historically Black community in the western United States, we designed an educational experience meant to foster American Black girls’ critical AI literacy and agency \cite{tissenbaum2021case}, such that they can thoughtfully reason about the nuances of AI systems and use them for expressive purposes that matter to them. 

\textbf{We investigated how fashion design could anchor a culturally responsive learning experience that supports Black girls’ critical engagement with and creative application of AI.} Our community partner organization suggested that fashion could be particularly felicitous to our goals because it sits at the intersection of identity, culture, self-expression, and creativity. Fashion serves as a powerful bridge between personal identity and learning. Incorporating fashion into STEM education offers potential for  students to see themselves and their cultures reflected, which fosters a stronger sense of belonging. When students feel their identities are acknowledged, they are more likely to engage deeply and take ownership of their learning \cite{saucier2011cape, singer2020foster}. For Black girls, whose identities are often stereotyped or misrepresented in digital spaces \cite{epps2020racism}, fashion could be an engaging context to critically investigate the shortcomings of AI, reflect on how AI intersects with societal power systems, and reimagine AI behavior that better centers a diverse range of stakeholders. In this work, we ask:

\textbf{RQ:} \textit{How are different facets of intersectional identities resources to foster critical AI literacy development through creative endeavors with AI?}

We analyze participation in, and artifacts from, a three-day workshop where five Black teen girls used GenAI tools to design fashion collections. We find that participants' creative explorations yielded opportunities for insightful critical reflection on the limitations of AI. Learners used their funds of knowledge to evaluate the AI behavior, as they engaged in creative endeavors. When the learners were positioned as agentic creators (i.e., where GenAI was a tool at their disposal, and they were the driving visionaries), they actively and thoughtfully accepted, rejected, critiqued, and developed strategies for shaping GenAI outputs. Our work contributes to computing education by demonstrating how culturally responsive creative learning challenges, like fashion design, can support critical AI literacy development and offer pathways for youth to meaningfully engage with, critique, and envision better versions of emerging AI technologies.

\section{Related Work}

\subsection{AI Literacy}
AI literacy is increasingly essential for youth to navigate, critique, and shape the AI systems that increasingly affect their lives. This literacy includes both technical (e.g., understanding AI algorithms, input-output behaviors, and machine learning mechanisms and limitations) and socio-ethical aspects (e.g, recognizing AI’s potential to reproduce problematic bias, influence decision-making, and affect equitable access to resources) \cite{Long2020, Touretzky2019, buddemeyer2022unwritten}.

AI literacy has been defined in multiple ways. For instance, Long \& Magerko (2020) define AI literacy as a set of competencies including understanding AI systems’ input–output behavior, learning processes, and limitations \cite{Long2020}, while Ng et al. (2021) outline our domains: knowing and understanding AI, using and applying AI, evaluating and creating AI, and contemplating AI ethics \cite{Ng2021ConceptualizingAL}. Approaches to fostering AI literacy among youth often emphasize data-driven, experimentation with and/or use of machine learning \cite{Morales-Navarro2024}. Many tools provide hands-on experiences with data and AI tools, including Scratch packages \cite{Jordan2021}, Machine Learning for Kids \cite{Lane2017MLKids}, and Teachable Machine \cite{Carney2020}.

While much existing work focuses on the technical aspects of AI literacy, recent literature emphasizes the importance of socio-ethical or critical AI literacy. This critical approach invites learners to interrogate who designs AI systems, which values they reflect, and who is impacted \cite{salac2023funds, solyst2025investigating, solyst2023potential, solyst2023would}. For example, prior work with Black girls (aged 9-12) supported their design and analysis of responsible AI \cite{solyst2023would}. In this work, young learners not only develop technical skills to understand AI systems but also engage critically with the social and ethical implications of AI. This emerging area of research aims to empower youth to move from users to co-creators, fostering deeper engagement, creativity, and a sense of ownership over the AI systems that impact their communities.

However, these recent efforts often take a structured support approach (e.g., \cite{salac2023funds, solyst2025investigating, solyst2023potential, solyst2023would, solyst2024children, solyst2023investigating}), which makes it unclear how  youth could draw upon their funds of knowledge in more open or self-directed contexts. For example, some work uses scenarios about unfair AI to encourage reflection \cite{salac2023funds, solyst2023would}. Other work encourages discussion of pre-selected examples of AI with problematic biases \cite{solyst2023potential} or prescribes structured ways of auditing (i.e., searching for and reporting) AI issues \cite{solyst2025investigating, morales2025learning, morales2024youth, solyst2025critical}. However, as we discuss in the next section, prior work on culturally responsive pedagogy, both in computing and beyond, suggests that designs can better center marginalized youth by providing more flexible structures (including open-ended projects) that allow learners to decide how best to use their funds of knowledge to critique existing systems and reimagine more just futures \cite{ladson1995toward, solyst2022running}.

\subsection{Culture and Computing Education}
Culturally Relevant Pedagogy (CRP) \cite{ladson1995toward} emphasizes how  instruction that draws on students' cultural identities, lived experiences, and community knowledge can support learners with a variety of backgrounds--not just those of socially dominant groups. CRP is a longstanding approach to promoting equity in education.  One approach to CRP theory we draw from is Lee's (2003) Cultural Modeling, which highlights three design principles: (1) Prior Knowledge and Cultural Models as Ways of Knowing (relating cultural practices and models to the academic context), (2) Engagement and Motivation (supporting  meta-cognition through interpersonal discourse), and (3) Social \& Civic Empowerment (by incorporating content that invites reflecting on community or personal needs)  \cite{lee2003toward}. Cultural Modeling was originally developed to support more traditional (textual and mathematical) literacies. As we elaborate below, these design principles are also relevant to supporting AI literacy. 

Scholars of Culturally Responsive Computing (CRC) apply CRP to computing education, showing how computing learning environments can affirm students’ identities and values, connect to their lived realities, and empower them as creators, not just the users of technology. This work has been conducted with marginalized learners of different backgrounds,  including Native American, Black students, and girls \cite{Eglash2006, scott2015culturally, li2025technology, solyst2022understanding}. CRC has three pillars: Asset-building (learning new skills in ways that build onto one's funds of knowledge), Reflection (reflection on one's own positionality and systems of power in computing technologies), and Connectedness (connection and responsibility to those in the learning experiences and broader communities) \cite{scott2015culturally}. Building on this work, Tanksley (2024) highlights how Black students use algorithmic literacies to challenge, rebuild, and re-imagine the algorithmic world \cite{tanksley2024changing}. Shaw et al. use womanist storytelling to reimagine Black girls’ participation in computing futures \cite{ShawMiaS2023RaBg}. Solyst et al. developed culturally responsive AI learning experiences for Black girls \cite{solyst2023would}. Coenraad \& Weintrop explored how Black youth engage critically with AI through the lens of “techQuity” (technology and equity) \cite{coenraad2024talking}. 

Scholars in the creative computing and maker education also emphasize importance of community-connected, identity-affirming design. Roque and Jain (2018) emphasize that young people flourish as creative computing facilitators when their cultural assets are recognized in out-of-school contexts \cite{roque2018becoming}. Vossoughi et al., 2021 argue for “pedagogies of joint activity” that move beyond binaries of adult- versus child-centered learning, offering a model of learning that is relational, participatory, and justice-oriented \cite{Vossoughi2021}. Erete et al. (2021) and Pinkard et al. (2020) advocate for transformative justice approaches girls of color in developing both computational skills and socio-political awareness \cite{Erete2021, pinkard2019equitable}. We see great opportunity to better understand and subsequently design critical AI literacy education with a culturally responsive lens.

\subsection{Opportunities for Fashion in STEM Learning}

Drawing on culturally responsive approaches to computing education, we saw facilitating open-ended creative context as particularly fitting to develop critical AI literacy. Fashion, in particular, reflects both individual identity and cultural heritage by incorporating social values and traditions \cite{davis1994fashion}. Historically, clothing has been used as an indication of one’s societal position and to identify oneself in a public space \cite{crane2000fashion, simpson2020youths}. More notably, fashion can signal group belonging and personal expression \cite{simpson2020youths, jenkinson2019wear}, such as in the Cape Verdean youth of Greater Boston who use hip-hop fashion to express their Blackness and creole cultural identities \cite{saucier2011cape}. Fashion is also a means of expressing one’s values, such as religion, with many religions having specific garments that hold symbolic significance \cite{shimek2012abaya, almila2020introduction}.

Taken together, fashion offers a unique way to motivate youth by connecting STEM education to personal identity and culture through creativity. As a form of art, fashion fits with the STEAM approach, which blends art and creativity with STEM learning to increase engagement \cite{bequette2012place}. For many youth, fashion serves as a relatable gateway to explore relevant STEM concepts \cite{ogle2019fashion, stewart2020promoting}. Prior work has shown that programs that combine fashion with STEM learning help deepen technical understanding while building a strong foundation for future academic and career success at this intersection \cite{dunne2015fashion, ogle2019fashion}. In fact, there has also been a recent push for fashion educators to consider applying social justice pedagogy into fashion education, involving participatory action research to increase social justice awareness \cite{reed2023creating}. In alignment with culturally responsive goals, fashion is ripe for exploration in fostering critical AI literacy, which has yet to be explored to our knowledge.

\section{Methods}
We ran an IRB-approved educational workshop study with a youth-centered community center in a predominantly Black neighborhood of a large urban city in the Pacific Northwest region of the United States. 

An administrative staff member from the community center led the outreach and recruitment efforts for the program, which focused on Black girls between the ages of 13 and 18. We selected this age range in part due to minimum age requirements associated with some of the AI technologies used in the program. Recruitment was conducted through local middle and high schools in the city's school district.

Our deliberate choice for working with teenage youth was due to the fact that at this age, youth are using AI-driven technology frequently. We also chose to specifically work with Black teenage girls, since women and people of color are still underrepresented as AI creators \cite{slaterintersectional}, resulting in facing heightened harms from AI but untapped potential to involve Black girls in broader AI ethics discourse. By chance, all youth who enrolled were also Muslim.

We worked closely with an administrative staff member to design the promotional flyer for the event, ensuring that it clearly communicated the program’s dual focus on fashion and the broader scope of GenAI. The flyer advertised that participants in the educational workshop study would receive  \$50 USD  in compensation.

Five Black girls, ranging in age from 14 to 18, were enrolled in the program. Table \ref{tab:participant_info} provides demographic details and a summary of participants’ initial knowledge relevant to the study.

\begin{table*}[ht!]
\centering
\begin{tabular}{|c|c|c|}
\hline
\textbf{Participant ID} & \textbf{Age} & \textbf{Prior Knowledge or Experience With AI} \\
\hline
P1 & 14 & Yes; heard about AI online in Google search results and TikTok \\ \hline
P2 & 14 & Yes; used and thought that TikTok had AI \\ \hline
P3 & 16 & Yes; used and thought that Tiktok and Youtube had AI \\ \hline
P4 & 17 & Yes; used GenAI to create stories and for inspiration \\ \hline
P5 & 18 & Yes; used tools like Google Translate and ChatGPT \\
\hline
\end{tabular}
\caption{Demographic and background information of study participants. All participants were Black/African/African American Muslim girls, all of whom also had some prior experience and awareness of AI-powered tools.}
\label{tab:participant_info}
\end{table*}

\subsection{Workshop Design}
The workshop was arranged into a 3-day, 4-hour per-day workshop during the Spring Break week of the local school district in the urban city. We first introduced the concepts of AI, bias, and fairness by asking open-endedly what their experiences were with AI (gauging prior knowledge). We then showed examples of common use cases of AI (search engine results, biometric identifiers, etc.) with biased outputs and asking learners to discuss those biases. We introduced the more technical concepts of algorithm (an input, set of steps, and output), as well as data and training data in a short lesson at the beginning of the workshop. We then focused on the following fashion-related activities for the rest of the workshop.


\subsubsection{AI or Real? Guessing Game}
\begin{figure}[htbp]
  \centering
  \includegraphics[width=\columnwidth]{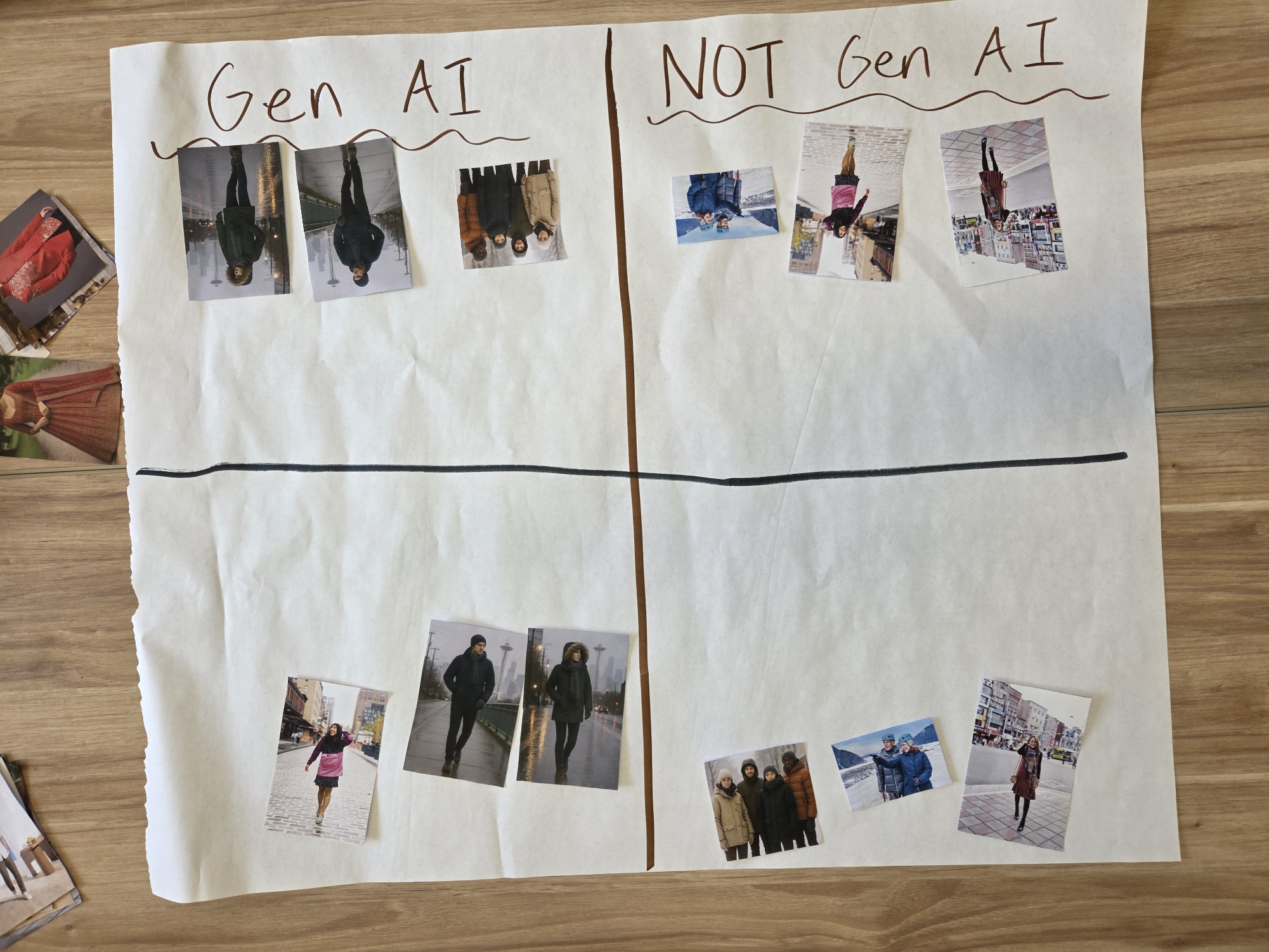}
  \caption{AI or Real? Guessing Game Activity}
  \label{fig:ai_or_real}
\end{figure}

We presented a guessing game in which several images were presented to the learners. Some were of human-ideated fashion designs, and others were generated by AI tools. The learners were asked to identify which designs appeared to be AI-generated and which were not (Figure \ref{fig:ai_or_real}). This provided initial insight into the potential patterns or limitations of GenAI designs.

\subsubsection{Can AI Represent Your Current Style?}
\begin{figure}[ht]
  \centering
  \begin{subfigure}[t]{0.3\columnwidth}
    \includegraphics[width=\linewidth]{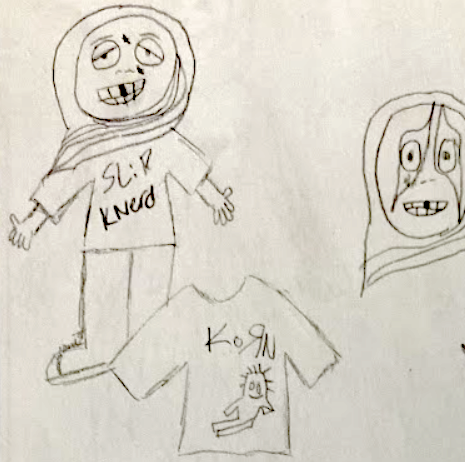}
    \caption{Hijabi wearing a KORN shirt}
    \label{fig:korn_shirt}
  \end{subfigure}
  \hfill
  \begin{subfigure}[t]{0.3\columnwidth}
    \includegraphics[width=\linewidth]{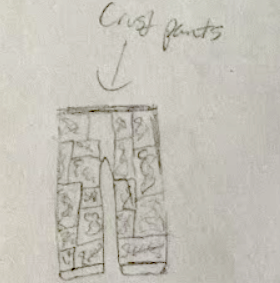}
    \caption{Crust pants \cite{crustpants2023}}
    \label{fig:crust_pants}
  \end{subfigure}
  \hfill
  \begin{subfigure}[t]{0.3\columnwidth}
    \includegraphics[width=\linewidth]{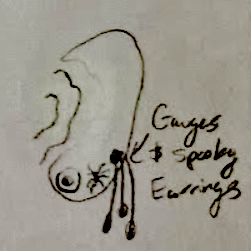}
    \caption{Gauges and spooky earrings}
    \label{fig:spooky_earrings}
  \end{subfigure}
  \caption{Outfits illustrated by P5 representing her personal fashion aesthetic (Vintage Romantic Punk)}
  \label{fig:P5_current_style}
\end{figure}

In this activity, learners reflected on their own style by drawing out ideas and descriptions on a worksheet (Figure \ref{fig:P5_current_style}). This activity had two main goals: helping learners explore their fashion and personal styles, and to give hands on-experience with generative AI output. The learners used various GenAI and prompted for images representing their personal style. They evaluated the results and discussed how GenAI did and did not represent their styles accurately. As they did this activity, they concurrently filled out a worksheet, which had three entries, where they wrote down the prompt they tried, whether or not they liked the output, and described their reasoning.

\subsubsection{Ideating Your Own New Fashion Line}
\begin{figure}[htbp]
  \centering
  \includegraphics[width=\columnwidth]{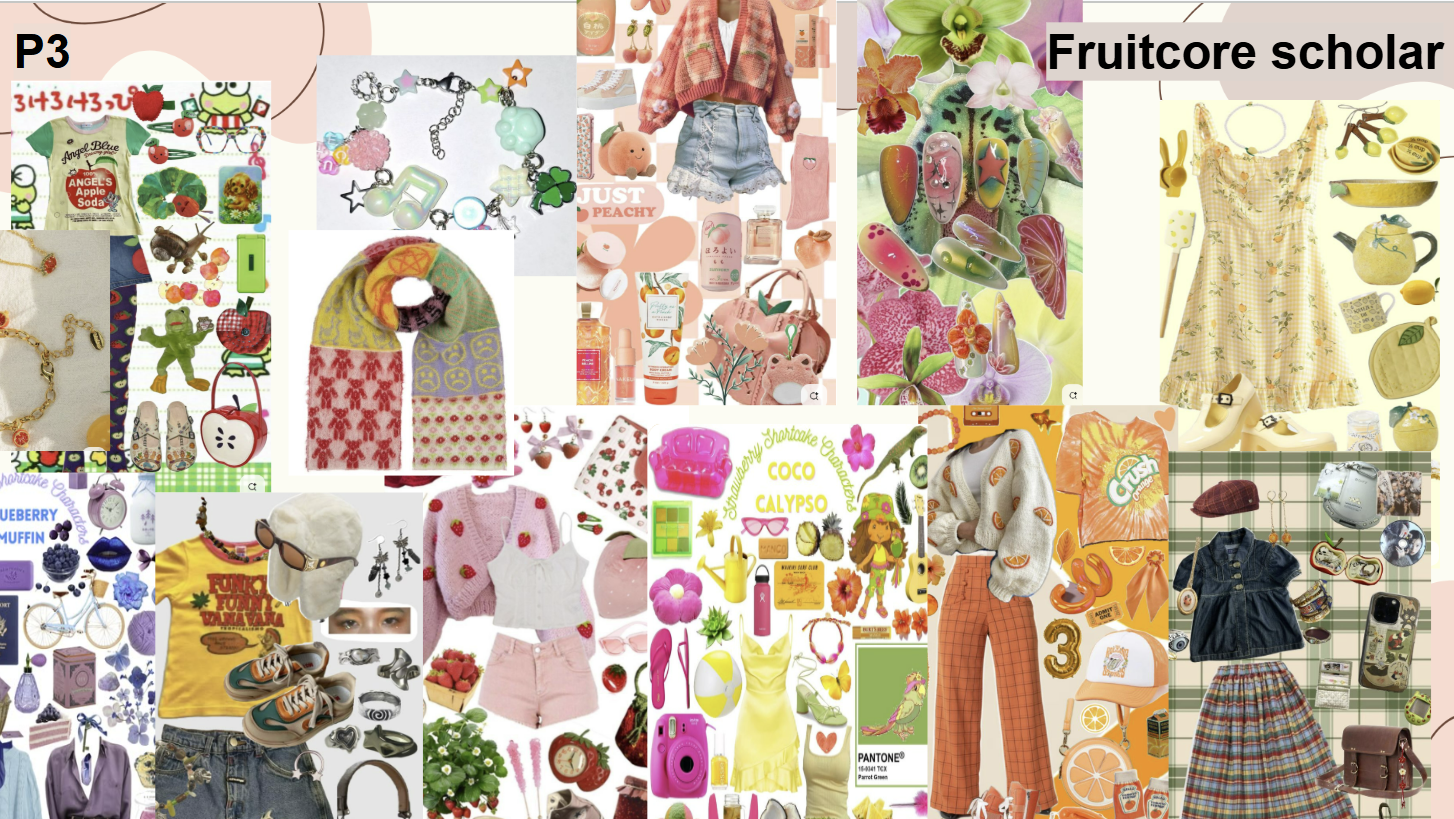}
  \caption{Example collage created by P3 for the ``Fruitcore Scholar'' fashion aesthetic.}
  \label{fig:fruitcore_collage}
\end{figure}

Learners were asked to invent a brand-new style or aesthetic. First, learners ideated and then collaged a new aesthetic, which was aimed at helping to develop a clearer visual understanding of their new fashion style, allowing them to better compare their expectations with the AI-generated results. Learners gathered images of outfits or clothing from various websites that reflected their chosen aesthetic. They compiled these images into a one-page digital collage to visually represent their style (Figure \ref{fig:fruitcore_collage}). They then used GenAI tools to bring their original designs to life, exploring how well the AI could interpret and visualize their creative ideas.

\subsubsection{Project Runway: Designing Your Own Fashion Collection}

Across the last two workshop days, learners created a fashion collection of five AI-generated outfits, based on the new aesthetics they had developed in the previous activity. Using GenAI (primarily OpenAI's ChatGPT version 4o) on individual laptops that we provided logged into a university account, they generated images as part of the collection with some additional types of outfits to create--sporty, celebration, and everyday—as part of a special design challenge.

This activity was designed to give learners the opportunity to reflect deeply on their experiences using GenAI in fashion. We encouraged them to think critically about how the AI tools worked and to make connections between their inputs and the outputs they received. Were the results aligned with their expectations? Did they encounter challenges while prompting the AI? Were there patterns or surprising elements, and why might those have occurred? Throughout their use of GenAI, they filled out the same worksheet as before, documenting their prompts and if they felt that the outputs were satisfactory and why.

After creating their collages, learners engaged in a group discussion about their prompting strategies, the quality of the outputs, and their thoughts on how fashion models should be represented in AI. They also shared suggestions for how the AI tools could be improved. Finally, everyone presented their designs to the group, and peers offered feedback or asked questions. Their peers in the audience highlighted one inspiring aspect and one question or area of curiosity for each design.

\subsection{Data Collection and Analysis}
The first author led the session and at least two other student researchers took notes during the session. Student researchers were trained and shown examples of notes to prepare them for the data collection. As notetakers, they wrote down their observations of the sessions, including transcribing conversations. They also took pictures of artifacts (e.g., worksheets) for later analysis. We captured audio data with microphones running through the entire session in case any transcriptions of conversations were incomplete. All screens for computers were also recorded while in use, so we could see how youth were interacting with GenAI. Lastly, we downloaded GenAI logs from their sessions. In total, we captured approximately 11 hours of screen recordings, 7 hours of audio recordings, and analyzed 131 generated images from the learners. Close analyses of the screen and audio recordings were done based on key moments indicated in the detailed notes.

For data analysis, we used consensus-based \cite{hammer2014confusing} inductive thematic analysis \cite{clarke2014thematic} on all sources of data from the sessions. The first three authors of the paper conducted close analysis of the data, coming up with initial codes and then grouping those into subthemes and then larger themes. Analysis was iteratively checked between these first three authors in feedback cycles with feedback from the last two authors.

\subsubsection{Researcher Positionality}
Our positionalities as researchers at a large university impact how we approached this work. For context, we describe them here. Our team includes adults with American, Christian, Muslim, and Jewish cultural backgrounds. Racially, our team includes Asian, White-presenting, and Black individuals. Some our team members have a long history of conducting design-based research with youth and new computational media, with training in human-computer interaction, information studies, computer science, learning sciences, and games. We are both motivated to carry out this work and often frustrated by how rarely youth voices, especially those of minoritized youth, are part of technology design processes.

\section{Results}
In this section, we present how youth engaged with GenAI tools to explore identity, express creativity, and navigate the limitations of AI systems. To contextualize our findings with respect to participants' prior knowledge,  all learners had some experience with AI but not all with GenAI. Experience varied, from having heard about AI and using AI-powered apps (e.g., TikTok) or GenAI. The older youth in the workshop tended to have more experience with AI. As we will show, youth drew on their cultural knowledge, personal aesthetics, and lived experiences to develop unique styles and envision novel fashion ideas. As they worked with GenAI, they encountered challenges related to misrepresentation and system bias, causing them to adapt their prompting strategies and reflect on how AI systems interpret identity-related input. In this section, we detail the following: (1) how youth used fashion as a domain for identity expression and innovation, (2) how they learned to iterate and adapt their GenAI prompts in response to system limitations, and (3) how these interactions supported emerging critical reflections on bias, representation, and the socio-ethical nature of GenAI.

\subsection{Fashion as a Creative Context for Identity and Expression}
Youth drew on their funds of knowledge and identities to envision and express creative fashion ideas with GenAI. They blended aspects of their cultures and identities, including \textit{U.S. teen culture, girl culture, African-American culture,} and \textit{Muslim culture} (all of which constituted aspects of their intersectional identities), to create garment designs that reflected and transcended traditions, embracing symbolic elements like color and accessories, and using GenAI tools to envision and create high-fidelity representations of their imaginative and personalized styles. In the following subsections, we characterize how participants used fashion as a creative domain for identity-linked expression and critical inquiry into GenAI system behavior. 

\subsubsection{Style Merged Cultural Values, Identity, and Fresh Trends}
To characterize how the youth engaged with style and fashion, we saw youth engage deeply with the reflective activities before using GenAI. When collaging and developing their styles, they combined traditional elements of religious clothing and fresh current styles. For example, participants alternated between traditional Muslim clothing (e.g., abayas and hijabs) and elements popular in Western-style outfits (e.g., hoodies or trendy items like \emph{``crust pants’’}). P5 explained, \emph{“My current style is a combination of a lot of traditional wear that I’ve been brought up with, and I also like to wear a lot of Western stuff… I want to incorporate my new styles into [traditional] things I already love.”} Learners emphasized that style was about \emph{“self expression”} (P3), as they mixed and matched elements of fashion.

P4 described how her style collage included a \emph{``Utero Nirvana shirt’’} because \emph{``Nirvana was one of the first few rock bands [she] was introduced to.’’} She concluded, \emph{``I’ve also seen a lot of people make a lot of streetwear with it, with their brand \& style, and I do like it... I have some pins and stuff from Nirvana, and I think that's dope.’’} P5 described mixing moody and trendy aesthetics with modesty: \emph{``To keep my traditional wear, I wanted to try jean skirts... I also wanted to do dark colors.''} These are but a few examples of how everyday fashion provided grounding for the girls' exploration, innovation, and expression of their intersectional identities in ways that upheld their cultural and aesthetic values.

\subsubsection{GenAI Enabled Imaginative and Complex Fashion Ideas}
In comparison to using image search engines for innovating styles, the youth realized that GenAI could create images of new innovative fashion ideas. For example, P5 noted how she \emph{``could not find image[s] of spiky scarves on Google, so using generative AI would be useful.’’} In various cases, like this, GenAI supported participants in dreaming up and bringing to life new ideas that inspired them. They came up with a variety of aesthetics (Figure \ref{tab:learners_aesthetics}).

\begin{table*}[ht!]
\centering
\begin{tabular}{|c|c|m{4in}|}
\hline
\textbf{Participant ID} & \textbf{New Aesthetic Name} & \textbf{Description} \\
\hline
P1 & Preppy Poet / Bows Extravaganza & \emph{``lighter pinks… preppy also means bows. Any clothes or accessories with bows.”} \\ \hline
P2 & Fairycore Pastel & \emph{``think of wings, think of light colors, think of very girly, light colors, flowers, bows.”} \\ \hline
P3 & Fruitcore Scholar & \emph{``centered around fruits… bring color to boring everyday outfits, the one that you work in or go to school in.”} \\ \hline
P4 & Modest Muse & \emph{``dark-themed streetwear… lots of black grillz, comes from Japanese culture… African Americans have been embracing that"} \\ \hline
P5 & Vintage Romantic Punk & \emph{"fun colors, love the spikes, patches on the pants, spiky jacket"} \\
\hline
\end{tabular}
\caption{Table of learners’ styles and summarizing quotes.}
\label{tab:learners_aesthetics}
\end{table*}

\begin{figure}[h]
  \centering
  \begin{subfigure}[t]{0.48\columnwidth}
    \includegraphics[width=\linewidth]{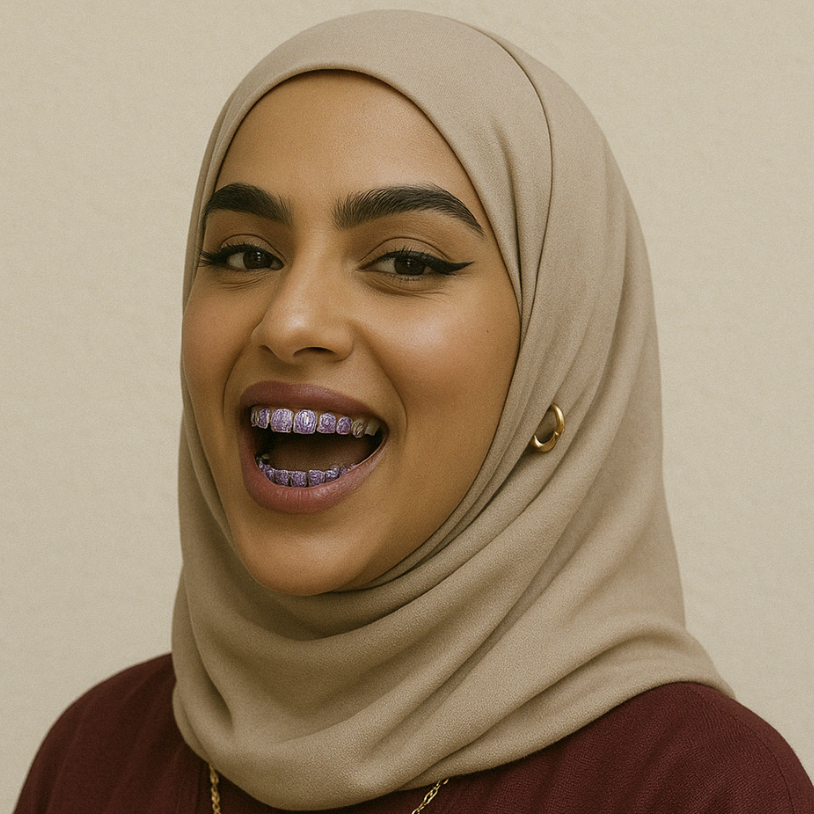}
    \caption{Hijabi wearing purple grillz (P4, Modest Muse)}
    \label{fig:purple_grillz}
  \end{subfigure}
  \hfill
  \begin{subfigure}[t]{0.48\columnwidth}
    \includegraphics[width=\linewidth]{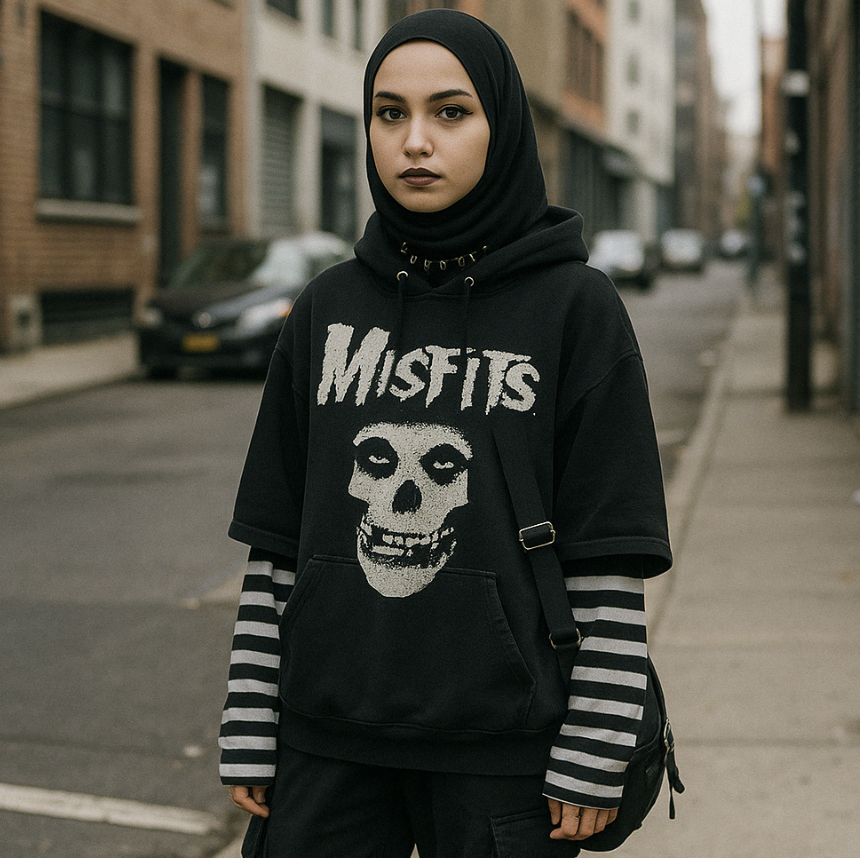}
    \caption{Hijabi wearing a Misfits shirt (P4, Modest Muse)}
    \label{fig:nirvana_shirt}
  \end{subfigure}

  \vspace{1mm}

  \begin{subfigure}[t]{0.48\columnwidth}
    \includegraphics[width=\linewidth]{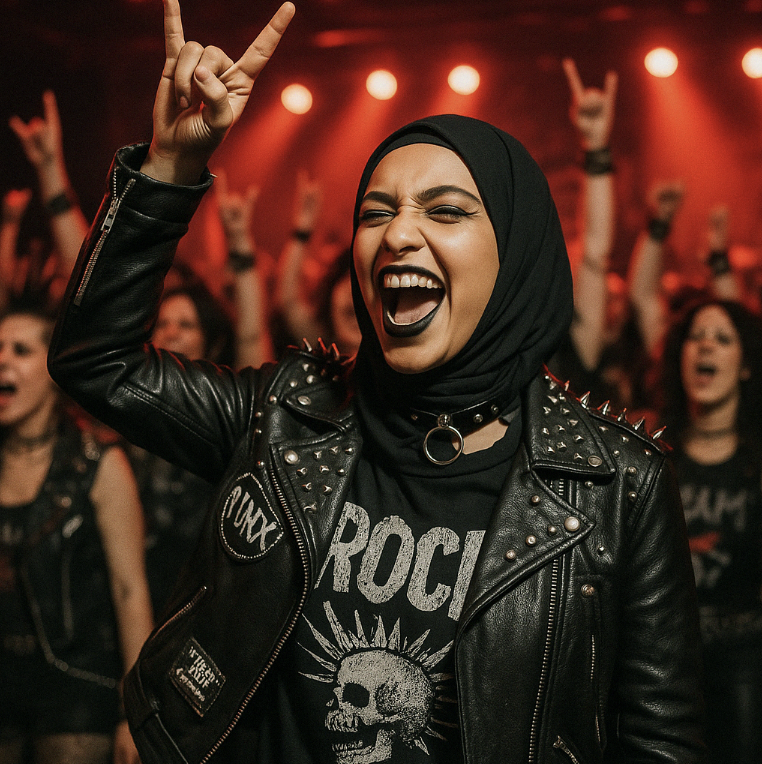}
    \caption{Hijabi wearing a spiky leather jacket (P5, Vintage Romantic Punk)}
    \label{fig:spiky_leather_jacket}
  \end{subfigure}
  \hfill
  \begin{subfigure}[t]{0.48\columnwidth}
    \includegraphics[width=\linewidth]{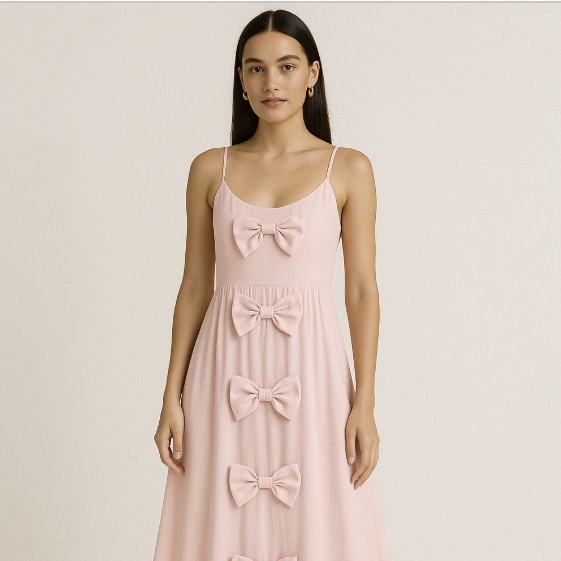}
    \caption{Model wearing a sleeveless light pink dress with bows. (P1, Bows Extravaganza)}
    \label{fig:pink_dress_bows}
  \end{subfigure}

  \caption{Examples of satisfactory fashion designs across multiple fashion aesthetics}
  \label{fig:satisfactory_designs}
\end{figure}

Ideas that were truly innovative, i.e., novel combinations of clothing garments and styles, were particularly exciting but also especially difficult to produce satisfactory representations of. For example, P5 went on to describe how her hybrid style blended grunge (band T-shirts) and modesty (hijab), since \emph{“you don’t see people wearing a lot of western stuff with scarves and hijabs.''} P4 also had a unique combination that she wanted to generate–a hijabi woman wearing black grillz; \emph{``black grillz aren't really that common or shown around a lot.’’} P4 went on to describe that it took quite a bit of prompt engineering–\emph{``I was trying to get grillz on a hijabi that is wearing a bunch of streetwear basically … I had to change up the wording a lot. It would often make her makeup really dark or give her black lipstick, so I had to specify and re-specify and say black teeth, black grillz, and finally got it, which is everything,’’} (Figure \ref{fig:purple_grillz}). The excitement was notable when the learners got the GenAI to output images that were well-aligned with their expectations. However, not all outputs were satisfactory.

\subsection{Learning to Prompt: Frustration, Iteration, and Adaptation to GenAI Constraints}

Learners assessed satisfactory outputs with their funds of knowledge and ideas based on their initial reflective activities, where they defined their style and collaged their new aesthetic using search engine results. As youth worked with GenAI tools, they changed how they prompted as they encountered the limitations of AI. They frequently revised their inputs to refine results but noted that tools often misinterpreted inputs, were insensitive to representing cultural garments, and required excessive detail.

\subsubsection{GenAI Misinterpretation of Inputs: Literal Outputs and Trouble with Trends}

\begin{figure}[ht]
  \centering
  \begin{subfigure}[t]{0.48\columnwidth}
    \includegraphics[width=\linewidth]{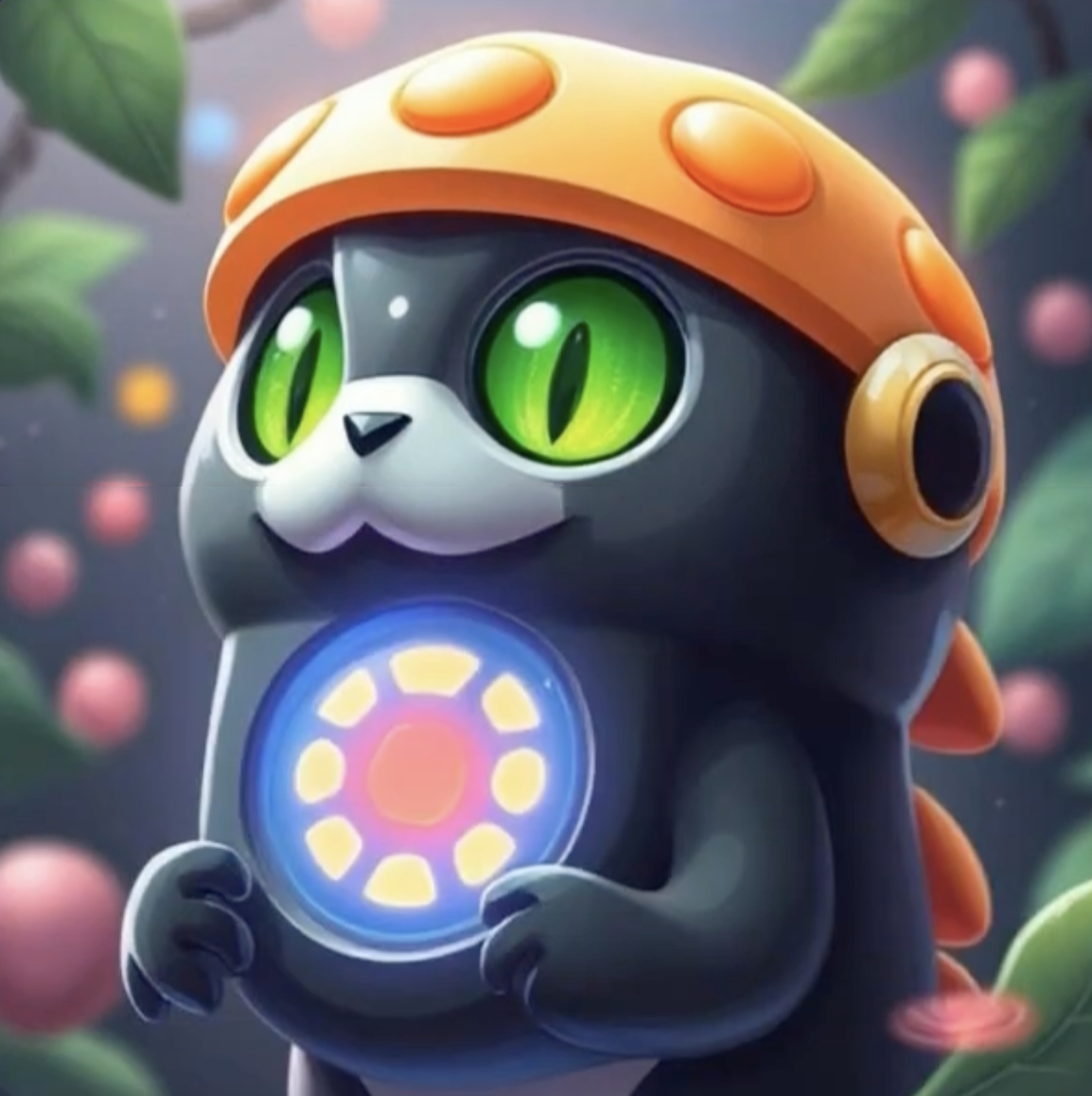}
    \caption{Incorrect interpretation of "Juminocore" \cite{juminocore2025}. (P3, Fruitcore Scholar)}
    \label{fig:juminocore}
  \end{subfigure}
  \hfill
  \begin{subfigure}[t]{0.48\columnwidth}
    \includegraphics[width=\linewidth]{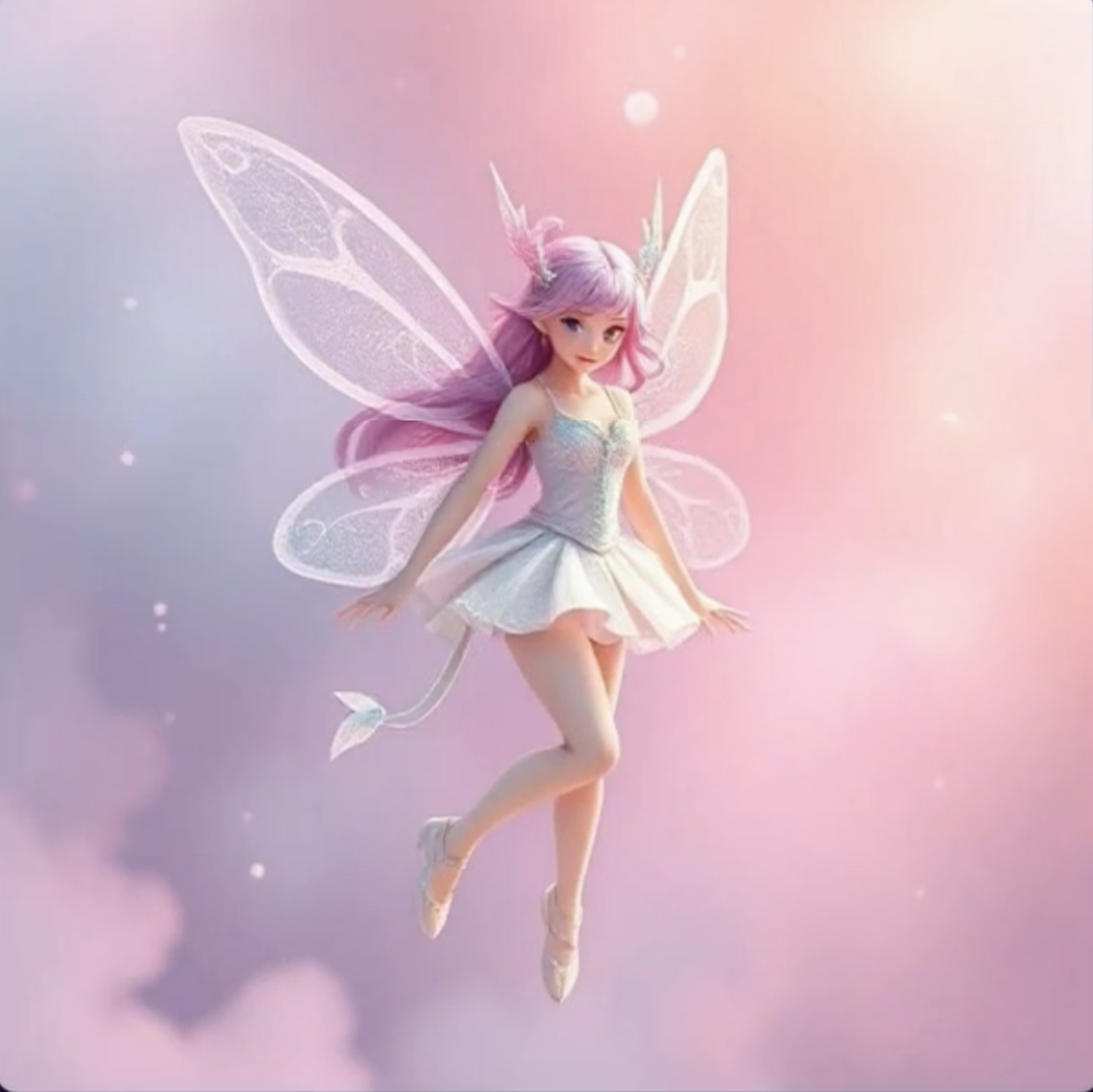}
    \caption{Image of a literal fairy (P2, Fairycore Pastel)}
    \label{fig:literal_fairy}
  \end{subfigure}

  \vspace{1mm}
  
  \begin{subfigure}[t]{0.48\columnwidth}
    \includegraphics[width=\linewidth]{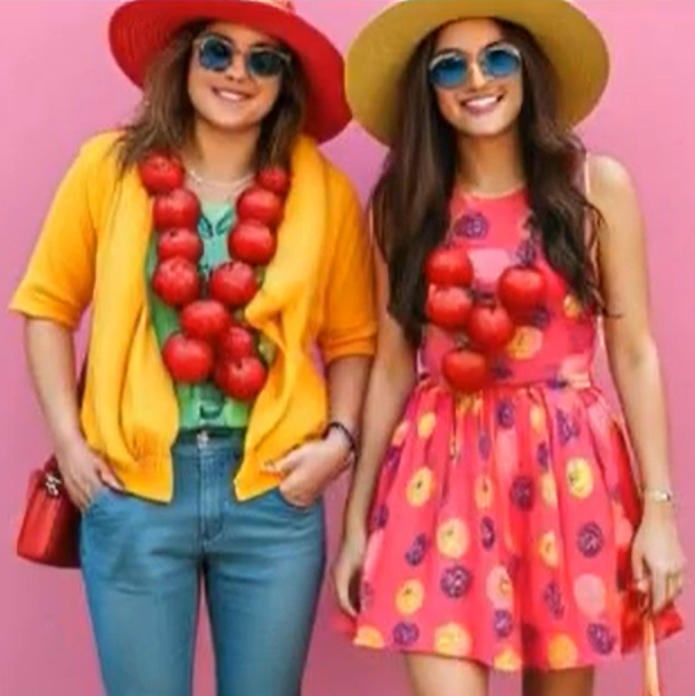}
    \caption{Image of outfits with literal fruits (P3, Fruitcore Scholar)}
    \label{fig:literal_fruitcore}
  \end{subfigure}
  \hfill
  \begin{subfigure}[t]{0.48\columnwidth}
    \includegraphics[width=\linewidth]{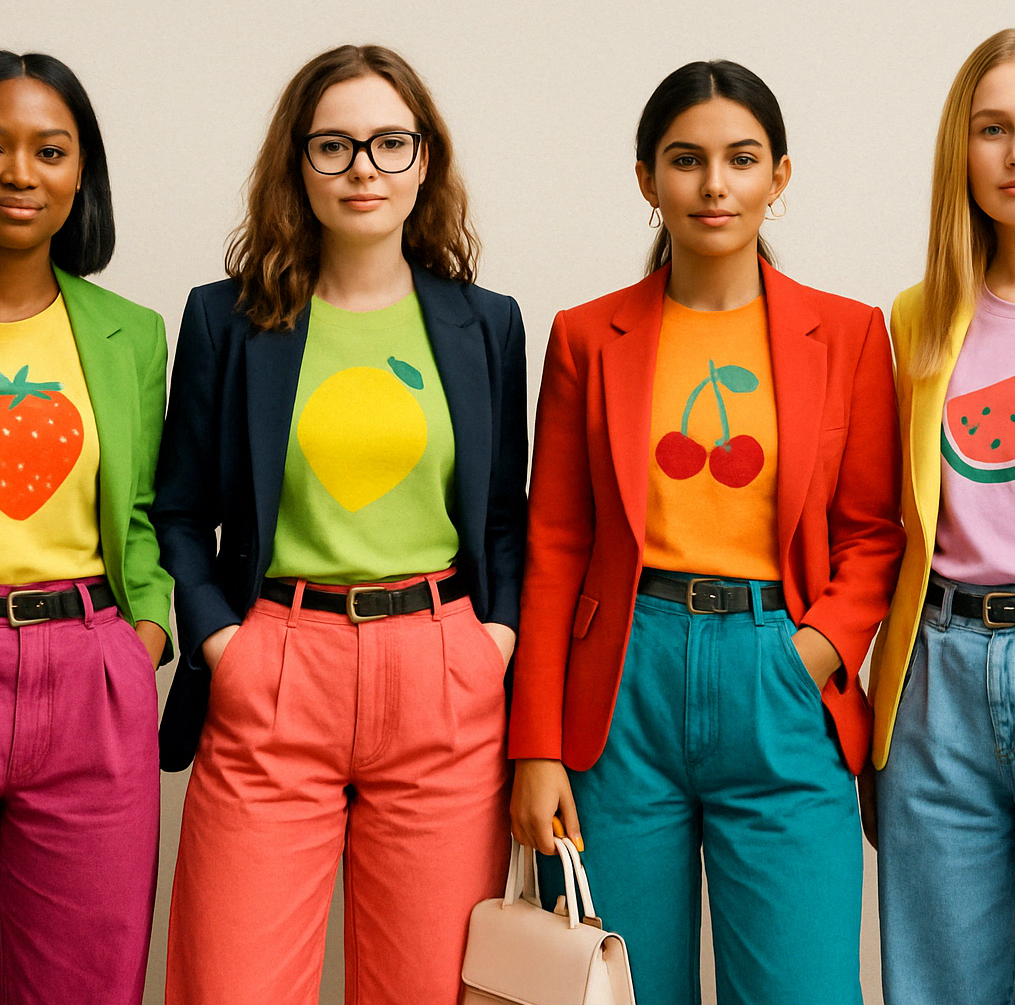}
    \caption{Image of preferred design for aesthetic (P3, Fruitcore Scholar)}
    \label{fig:preferred_fruitcore}
  \end{subfigure}
  \caption{Images of Fairycore and Fruitcore designs}
  \label{fig:bad_outputs}
\end{figure}

Throughout their use of GenAI, the youth pointed out how the AI did not understand their inputs. Sometimes the outputs were very literal or lacked critical inference. P2, whose aesthetic was \emph{``pastel fairy core,’’} found that the AI often generated literal images of fairies (Figure \ref{fig:literal_fairy}). P3, who made the fruitcore scholar collection, noted that she was \emph{``getting the same thing–they just keep generating a picture of fruit on ChatGPT and call[ing] it a day’’} (Figure \ref{fig:literal_fruitcore}). Instead, she would rather have smaller fruit prints that more strongly emphasized the diversity of colors in the \textit{likeness} of fruits (Figure \ref{fig:preferred_fruitcore}).

Unsatisfactory results were especially the case when the GenAI did not interpret trendy vocabulary and niche fashion references. For example, when P3 tried using an existing yet less well-known aesthetic, \emph{``Juminocore,\footnote{According to the Personal Aesthetics Wiki \cite{juminocore2025} Juminocore is a heavily Japanese retro inspired aesthetic popular on Pinterest and Tiktok.}''} directly in her prompts, ChatGPT did not correctly generate the fashion trend (Figure \ref{fig:juminocore}). Rather, perhaps because the term originated from Japanese, the system output Japanese-like images, but they did not include the right content. We saw this as a creative barrier to innovation: when young people are on the cutting edge of experimenting with new trends and ideas, GenAI systems are not necessarily well-equipped to support that innovation.

\subsubsection{Youths’ Prompting Shifted to Accommodate GenAI Limitations}
Through our analyses of observations and logs of youths’ iterative (re-)prompting, we saw that youth progressively reworded their prompts, increasing specificity and length as they tried to improve GenAI outputs. They found this frustrating: P1 described that the outputs \emph{``are not that bad, but ChatGPT is annoying because you have to like type out every single detail... If you forget a detail, you have to add it in.’’} P5 added that, \emph{``You can never be too specific. It took like twelve words to get the correct outfit.’’} The learners found that they had to translate different trends and terms that they were familiar with to descriptive language toward a White Western interpretation. For example, When P4 was trying to generate Black grillz, she noted how difficult it was to get Black grillz, saying how she had tried various terms: \emph{``Black teeth, black grillz, black this, black that.’’}

\begin{figure}[ht]
  \centering
  \begin{subfigure}[t]{0.3\columnwidth}
    \includegraphics[width=\linewidth]{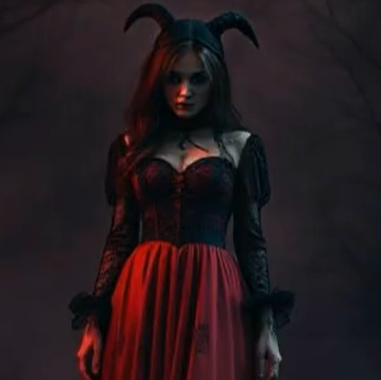}
    \caption{Image prompt: ``long red gothic spooky web dress''}
    \label{fig:red_dress1}
  \end{subfigure}
  \hfill
  \begin{subfigure}[t]{0.3\columnwidth}
    \includegraphics[width=\linewidth]{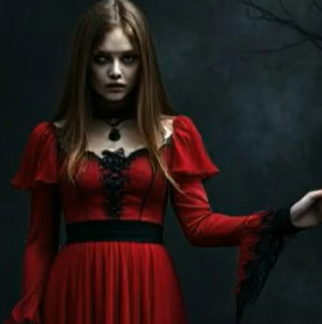}
    \caption{Image prompt: ``long red gothic spooky web dress modest''}
    \label{fig:red_dress2}
  \end{subfigure}
  \hfill
  \begin{subfigure}[t]{0.3\columnwidth}
    \includegraphics[width=\linewidth]{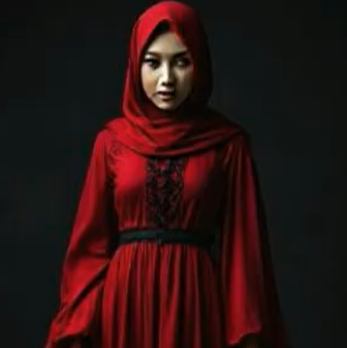}
    \caption{Image prompt: ``long red gothic spooky web dress modest hijab''}
    \label{fig:red_dress3}
  \end{subfigure}
  \caption{Several images from P5's iterative prompts for a ``gothic dress''}
  \label{fig:red_dress_prompts}
\end{figure}

Additionally, while we note that their prompting became more specific, it was especially the case that they had to prompt specifically for certain types of diversity and representation. For instance, P5 wanted to generate an image of a red gothic dress. She began with the prompt \emph{``gothic dress,''} followed by \emph{``long red gothic spooky web dress''} (Figure \ref{fig:red_dress1}) to incorporate more aesthetic details. The results were still not modest enough, so she modified her prompt to include \emph{``long red gothic spooky web dress modest''} (Figure \ref{fig:red_dress2}) and eventually \emph{``long red gothic spooky web dress modest poc''} in an attempt to generate an image featuring a modest version of the outfit worn by model of color, since all the models so far had been light-skinned. Finally she settled on \emph{``long red gothic spooky web dress modest hijab''} (Figure \ref{fig:red_dress3}), with the term \emph{hijab} being used to successfully generate an image of a modest red dress worn by a hijabi.

P4 brought up that she had to \emph{``address what are the most important details, like brown-skinned hijabi girls... will [ideally] generate girls close to that skin tone.’’} This, along with facilitator support, led the youth to realize and question ethical issues with GenAI, which we share more insights on next.

\begin{figure}[h]
  \centering
  \begin{subfigure}[t]{0.48\columnwidth}
    \includegraphics[width=\linewidth]{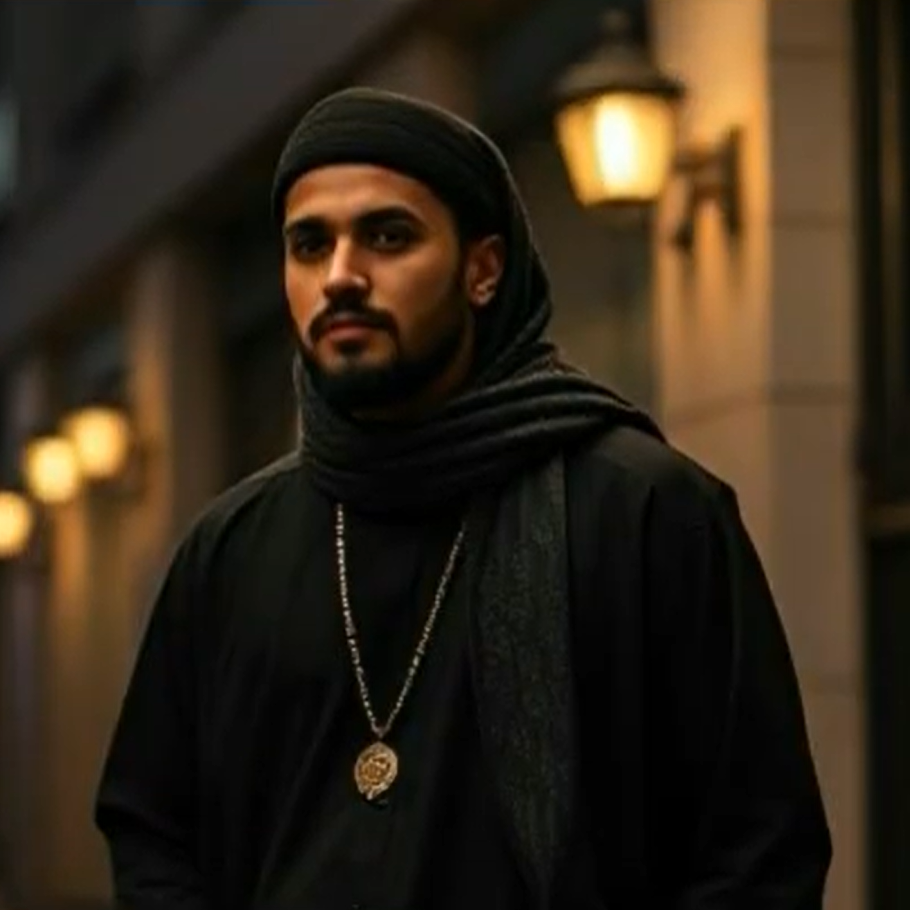}
    \caption{AI generated `Khaleeji' as a man (P4, Modest Muse)}
    \label{fig:khaleeji_man}
  \end{subfigure}
  \hfill
  \begin{subfigure}[t]{0.48\columnwidth}
    \includegraphics[width=\linewidth]{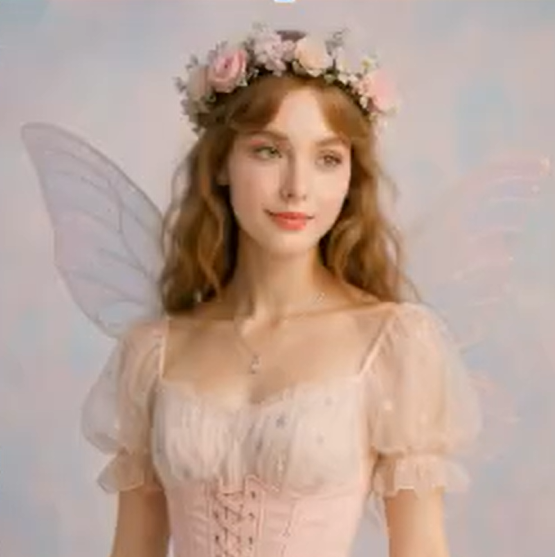}
    \caption{Model's default skin color as white (P2, Fairycore Pastel)}
    \label{fig:white_skinned_fairy}
  \end{subfigure}

  \vspace{1mm}

  \begin{subfigure}[t]{0.48\columnwidth}
    \includegraphics[width=\linewidth]{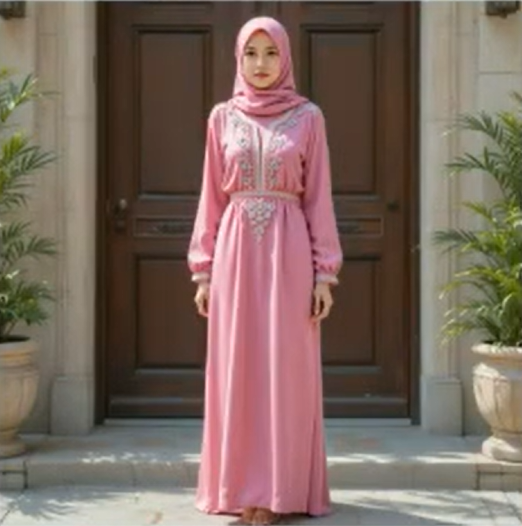}
    \caption{Inappropriate design for an abaya (P1, Bows Extravaganza)}
    \label{fig:abaya}
  \end{subfigure}
  \hfill
  \begin{subfigure}[t]{0.48\columnwidth}
    \includegraphics[width=\linewidth]{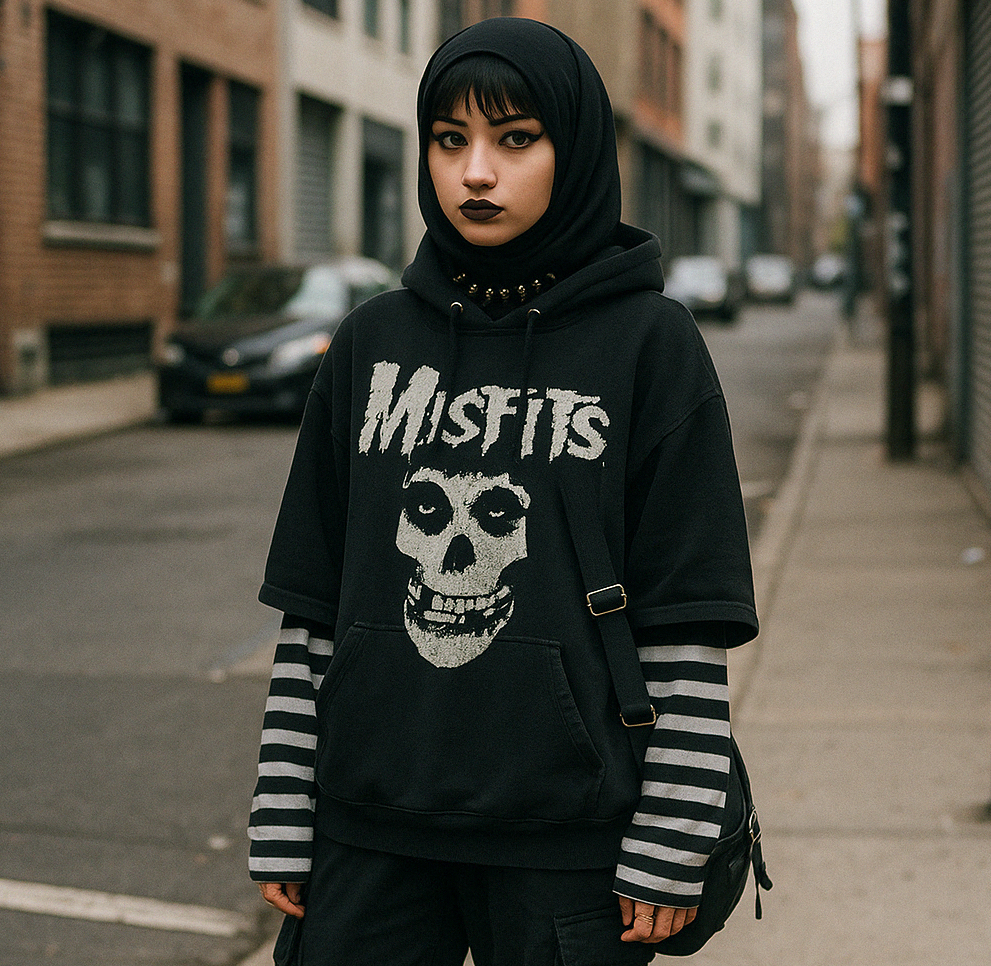}
    \caption{Hijab with bangs showing (P4, Modest Muse)}
    \label{fig:hijabi_with_bangs}
  \end{subfigure}

  \caption{Examples of unsatisfactory GenAI outputs}
  \label{fig:unsatisfactory_designs}
\end{figure}

\subsection{Emerging Critical AI Literacy: Bias, Representation, and Ownership}

Through their engagements, youth developed and shared critical insights about bias and representation in GenAI. They recognized the limitations of training data and questioned the fairness of generated outputs. In particular, we saw the youth use their intersectional identities as assets to assess the GenAI outputs.

\subsubsection{Lack of Diversity and Connections to Technical Aspects of AI}
One of the main issues that youth identified with GenAI outputs was the lack of diversity; most generated images of women had light skin and thin body types. P3 reported, \emph{``Most of the people I’ve generated are just White… I didn’t say what the person… is gonna look like, but… the default seems to be a White person.’’} P5 concurred, \emph{``I tweaked my search so many different times, but it was consistently showing white skin tones. It’s used to the majority of people that wear that style, like goth. [But] even when asking for a hijab, it keeps showing people with white skin—it doesn’t think about the other people that also fit the category.’’} When asked why they thought so many White models were being generated (Figure \ref{fig:white_skinned_fairy}), the learners were also able to connect the idea of lacking diversity to the concept of data and real-world societal biases. In other words, they were able to apply the lesson content from days prior to what they experienced with GenAI. For example, P4 suggested that \emph{``White girls dominate the [modeling] field, so that is all the AI knows,’’} and P3 added that \emph{``White girls probably make up most of the data used for the tools.’’} P3 went on to connect how the people who create technology may be a cause of AI bias, suggesting that bias was \emph{``not the AI’s fault. It’s on the people who gather the data and give it to the AI. They have their own bias that they are not aware of, and that is what causes bias.’’} P5 also commented that this was a problematic oversight, surmising that \emph{``maybe the developers [were] in a rush and lazy.’’} Their recognition of bias supported connection to technical aspects of AI, which they learned about earlier. We see this reflection on bias as key to critical AI literacy, as recognition of bias means that they understand the AI can reflect and amplify societal biases as a socio-technical system.

\subsubsection{Evaluation and Critique of Misrepresentation}
The youth used their intellectual and cultural assets, including knowledge of teen culture and different facets of identity including Black American, Muslim, and Black girlhood to assess and critique misrepresentation in the GenAI outputs.

The girls tried to generate various fashion aesthetics that included pieces of Muslim garments that wore modestly, including abayas and hijabs, as well as reflected regional cultures, such as Khaleeji. However, these terms often led to outputs that the girls disapproved of. For example, when the girls generated images of hijabs, the AI tools often did not output representations that upheld standards of modesty. P2 brought up how \emph{``with a hijab, you aren’t supposed to show hair, but the output ended up showing bangs,''} as shown in Figure \ref{fig:hijabi_with_bangs}. Abayas, too, often were generated such that more skin was showing than was customary or were overly ornate (Figure \ref{fig:abaya}). P5 noted how ChatGPT outputs \emph{``kept showing the chest in the dress. It was very frustrating [when it generated] a sheer top.’’}

Gender bias was also prominent. For example, the term Khaleeji was gendered, with P4 reporting that \emph{``every time [she prompted] `Khaleeji,' [the image was of] a man’’} (as shown in Figure \ref{fig:khaleeji_man}), despite the Arabic term referring to a person \emph{of the gulf} that could be of any gender. This was also the case when it came to makeup. P4 noted that, \emph{``People in my culture would not wear this [makeup]. It would seem very extremely rare to see someone wearing this… someone with dark make up, so I had to request it not to include that.’’} However, when she prompted \emph{``no makeup,’’} she found that \emph{``it [then] made the model look masculine,’’} replacing the whole face (with one she did not like as much) rather than just removing the makeup and keeping other elements of the image constant. 

\section{Discussion}
Our findings show how teens used GenAI tools to express their intersectional identities, critique algorithmic limitations, and develop emerging forms of critical AI literacy through a culturally responsive, creative activity. Drawing from their rich funds of knowledge and aesthetic values, youth used GenAI to create imaginative fashion styles that reflected and reimagined their intersectional identities. However, their creative processes were often disrupted by GenAI’s limitations, particularly in its struggles to interpret certain cultural terms, aesthetic trends, and diverse representations. Through iterative prompting and collaboration, participants adapted their strategies to account for GenAI's constraints, learning to ``work around’’ its defaults and biases, especially those skewed toward Whiteness, Western norms, and normative beauty standards. In doing so, youth developed an awareness of the socio-ethical underpinnings of AI bias, including the role of training data and developer assumptions, and made connections to broader questions of biased representation and fairness.

This work aligns with the ample evidence that shows meaningful learning happens in  environments that value and foreground youth interests and lived experiences \cite{ito2013connected,gonzalez2007funds, moll2006funds}. The funds of knowledge approach, for example, posits that learning environments should identify and incorporate learners’ skills, knowledge, and competencies when thinking about designing meaningful learning experiences. The open-ended and interest-driven structure of the GenAI fashion activities presented in this work was central to fostering youth creativity, self-motivation, and meaningful engagement. Moreover, expansive learning environments support identity exploration through personally and culturally relevant inquiry \cite{vakil2022youth}; our program design positioned youth as the experts of their own experiences. Participants chose their own aesthetic directions, drew from their lived cultures, and set personal design goals, which supported sustained engagement and self-directed learning. In particular, our work showcases how teens’ prior knowledge about the cultural nuances of fashion can be a key resource for their learning, including shaping their critical perceptions of AI. Others observed that by making space for students’ funds of knowledge in a pedagogical approach intended to elicit ethical thinking about technologies, students can draw from their own experiences in their ethical sensemaking \cite{landesman2024integrating}. We contribute to this existing corpus of knowledge, and argue that the intentional integration of students’ lived experiences and genuine interests is crucial if we are to provide them the support they need to develop a critical and socio-ethical understanding of AI tools.

The workshop design was inspired by Carol Lee’s Cultural Modeling framework, which centers learners’ everyday knowledge and cultural practices as intellectual resources for meaningful learning \cite{lee2003toward}. The design successfully created space for youth to bring rich cultural insights into the fashion design activities, such as knowledge of Black girlhood, Muslim fashion, internet aesthetics, and teen vernacular. Rather than treating these as peripheral, our approach positioned them as central to the learning experience. For example, participants’ use of terms like \emph{``crust pants,’’} \emph{``fairycore,’’} and \emph{``Khaleeji’’} were not just stylistic choices but sociocultural knowledge. However, the GenAI systems often failed to recognize or accurately render these terms, requiring the teens to translate their prompts to dominant cultural codes in order to generate satisfactory outputs. Ultimately, while GenAI enabled creative exploration, the teens experienced how the AI fell short when it came to their sociocultural knowledge. Overall, we found the three cultural modeling design principles to be especially effective. Understood through the lens of  cultural modeling, it was through these moments of struggle and critique of the systems, where youth used their funds of knowledge (i.e., Design Principle 1: ``Prior Knowledge and Cultural Models as Ways of Knowing’’) to evaluate and engage with GenAI, that critical AI literacy was fostered. Leveraging Design Principle 2: ``Engagement and Motivation,’’ we facilitated the youth to discuss with one another through interactive learning activities and then engage in broader meta-cognitive discourse together in their fashion collection presentations. This discourse supported learners in admiring each others’ creative work, while talking about the challenges of using GenAI and how it did not always serve their creative endeavors. Lastly, for Design Principle 3: ``Social and Civic Empowerment,’’ we created tasks where the youth would encounter the problematic biased limitations of GenAI and grapple with its nuances. We urge future designers and practitioners who are thinking about supporting youths' learning about and with AI tools to consider these principles, as well as the effectiveness of interest-driven learning as shown in this study and others before it. As youth develop a socio-ethical awareness of AI's limitations and positioning within society, we see an opportunity for the adults in their lives to scaffold youth to critique the AI tools in their lives as a method of learning about their capabilities and limitations, with the possibility of exploring their own perceptions about what value these tools bring to their lives. 

\subsection{Future Directions}
We see various avenues for future research and design. Future work could investigate how youth’s critical AI literacies evolve over time and in different creative landscapes beyond fashion, such as art, music, or civic media. For example, what affordances do different types of creative environments lend? Taking into account youths’ experiences in the study, future design work might explore tools that scaffold iterative prompting, support more accurate cultural representation, and foster critical reflection. Further, there’s a need for GenAI tools that can be meaningfully altered by youth, such that they work for the users at hand. Ultimately, this research contributes to efforts to reimagine AI literacy and design.

\section{Conclusion}
In this work, we investigated how teens used facets of their identity to foster critical AI literacy in a creative context. By drawing from their own cultural knowledge, aesthetic values, and lived experiences, participants not only explored innovative and personally meaningful frontiers in fashion design but also surfaced limitations and biases embedded in GenAI systems. Their iterative prompting strategies and critiques supported their development of critical AI literacy, as they learned to navigate and question the limitations of GenAI tools. This work contributes insights on: (1) how culturally responsive, creative contexts like fashion can serve as powerful entry points for youth engagement with GenAI; (2) how critical AI literacy can emerge through leveraging sociocultural knowledge from intersectional identity; and (3) design implications for creating AI literacy experiences that center diverse youths’ ways of knowing.

\begin{acks}
We want to thank the incredible youth participants who were in the study, our fantastic community partner: the William Grose Center for Cultural Innovation, who we had the honor to work with and learn from, and the University of Washington staff who supported logistics. We are grateful to the additional research assistants who supported data collection: Mike Deng, Calvin Tsai, Sai Sunku, and Dhruv Bansal. We also thank Thomas M. Philip for being an outstanding and formative thought partner. 
\end{acks}

\bibliographystyle{ACM-Reference-Format}
\bibliography{refs}

\end{document}